\documentstyle[amstex]{article}

\catcode`\@=11
\def\numberbysection{\@addtoreset{equation}{section}
        \def\theequation{\thesection.\arabic{equation}}}
\numberbysection

\begin{document}

\newlength{\lno} \lno1.5cm \newlength{\len} \len=\textwidth%
\addtolength{\len}{-\lno}

\setcounter{page}{0}

\baselineskip7mm \renewcommand{\thefootnote}{\fnsymbol{footnote}} \newpage %
\setcounter{page}{0}

\begin{titlepage}     
\vspace{0.5cm}
\begin{center}
{\Large\bf  $sl(2|1)^{(2) }$ Gaudin Magnet and its associated Knizhnik-Zamolodchikov equation}\\
\vspace{1cm}
{\large  V. Kurak $^{\dag }$\hspace{.5cm} and \hspace{.5cm} A. Lima-Santos$^{\ddag}$ } \\
\vspace{1cm}
$^{\dag}${\large \em Universidade de S\~ao Paulo, Instituto de F\'{\i}sica \\
Caixa Postal 66318, CEP 05315-970~~S\~ao Paulo -SP, Brasil}\\
\vspace{.5cm}
$^{\ddag}${\large \em Universidade Federal de S\~ao Carlos, Departamento de F\'{\i}sica \\
Caixa Postal 676, CEP 13569-905~~S\~ao Carlos, Brasil}\\
\end{center}
\vspace{1.2cm}
\begin{abstract}
The semiclassical limit of the algebraic Bethe Ansatz  method is used to solve the 
theory of Gaudin models for the $sl(2|1)^{(2)}$ R-matrix. We find the spectra
and eigenvectors of the $N-1$ independents Gaudin Hamiltonians .
We also use the off-shell Bethe Ansatz method to show how the off-shell Gaudin equation 
solves the associated trigonometric system of Knizhnik-Zamolodchikov equations.
\end{abstract}
\vspace{2cm}
\begin{center}
\end{center}
\vfill
\begin{center}
\small{\today}
\end{center}
\end{titlepage}

\baselineskip6mm

\newpage{}

\section{Introduction}

Integrable models of classical statistical mechanics \cite{Baxter, KBI} and
of two-dimensional quantum field theories \cite{Abdalla}, have a central
object: the ${\cal R}$-matrix ${\cal R}(u)$, where $u$ is a spectral
parameter, acting on the tensor product $V\otimes V$ of a given vector space 
$V$ and being a solution of the Yang-Baxter ({\small YB}) equation 
\begin{equation}
{\cal R}_{12}(u){\cal R}_{13}(u+v){\cal R}_{23}(v)={\cal R}_{23}(v){\cal R}%
_{13}(u+v){\cal R}_{12}(u),  \label{int.1}
\end{equation}%
in $V^{1}\otimes V^{2}\otimes V^{3}$, where ${\cal R}_{12}={\cal R}\otimes 1$%
, ${\cal R}_{23}=1\otimes {\cal R}$, etc.

The solution ${\cal R}(u)$ of Eq.(\ref{int.1}) is said to be semiclassical
if ${\cal R}(u)$ also depends on an additional parameter $\eta $ in such a
way that%
\begin{equation}
{\cal R}(u,\eta )=1+\eta \ r(u)+{\rm o}(\eta ^{2}),  \label{int.1b}
\end{equation}%
where $1$ is the identity operator on the space $V\otimes V$. The
\textquotedblleft classical $r$-matrix\textquotedblright\ obeys the equation 
\begin{equation}
\lbrack r_{12}(u),r_{13}(u+v)+r_{23}(v)]+[r_{13}(u+v),r_{23}(v)]=0.
\label{int.2}
\end{equation}%
This equation, called the classical Yang-Baxter equation, plays an important
role in the theory of classical completely integrable systems \cite{Semenov}.

Nondegenerate solutions of (\ref{int.2}) in the tensor product of two copies
of a simple Lie algebra {\rm g} , $r_{ij}(u)\in {\rm g}_{i}\otimes {\rm g}%
_{j}$ , $i,j=1,2,3$, were classified by Belavin and Drinfeld \cite{BD}.

The classical {\small YB} equation has an interplay with conformal field
theory, shortly described in the following way: in the skew-symmetric case $%
r_{ji}(-u)+r_{ij}(u)=0$, it is the compatibility condition for the system of
linear differential equations 
\begin{equation}
\kappa \frac{\partial \Psi (z_{1},...,z_{N})}{\partial z_{i}}=\sum_{j\neq
i}r_{ij}(z_{i}-z_{j})\Psi (z_{1},...,z_{N})  \label{int.3}
\end{equation}%
in $N$ complex variables $z_{1},...,z_{N}$ for vector-valued functions $\Psi 
$ with values in the tensor space $V=V^{1}\otimes \cdots \otimes V^{N}$ . $%
\kappa $ is a coupling constant.

In the rational case \cite{BD}, very simple skew-symmetric solutions are
known: $r(u)={\rm C}_{2}/u$, where ${\rm C}_{2}\in {\rm g}\otimes {\rm g}$
is a symmetric invariant tensor of a finite dimensional Lie algebra ${\rm g}$
acting on a representation space $V$. The above system of linear
differential equations (\ref{int.3}) is known \ as the
Knizhnik-Zamolodchikov ({\small KZ}) system of equations for the conformal
blocks of the Wess-Zumino-Novikov-Witten ({\small WZNW}) models of the
conformal field theory on the sphere \cite{KZ}.

The algebraic Bethe Ansatz \cite{FT} was formulated in a parallel reasoning
to the representation theory of the classical Lie algebras and it is a
powerful method in the analysis of integrable models. Besides describing the
spectra of quantum integrable systems, the algebraic Bethe Ansatz is also
used to construct exact and manageable expressions for correlation functions 
\cite{KBI}. Various representations of correlators were found by Korepin 
\cite{KO}, using this method.

The Babujian and Flume work \cite{BAF} unveils a link between the algebraic
Bethe Ansatz for the theory of the Gaudin models\cite{GA} and the conformal
field theory of WZWN models. In their approach, the wave vectors of the
Bethe Ansatz equation for an inhomogeneous lattice model render, in the
semiclassical limit, solutions of the {\small KZ} equation for the case of
simple Lie algebras. For instance, in the $su(2)$ example, the algebraic
quantum inverse scattering method \cite{FT} allows one to write the
following equation 
\begin{equation}
\tau (u|z)\Phi (u_{1,\cdots ,}u_{p})=\Lambda (u,u_{1},\cdots ,u_{p}|z)\Phi
(u_{1},\cdots ,u_{p})-\sum_{\alpha =1}^{p}\frac{{\cal F}_{\alpha }\Phi
^{\alpha }}{u-u_{\alpha }}.  \label{int.4}
\end{equation}%
Here $\tau (u|z)$ denotes the transfer matrix of the rational vertex model
in an inhomogeneous lattice acting on an $N$-fold tensor product of $su(2)$
representation spaces. $\Phi ^{\alpha }$ \ meaning $\Phi ^{\alpha }=\Phi
(u_{1},\cdots u_{\alpha -1},u,u_{\alpha +1},...,u_{p})$ ; ${\cal F}_{\alpha
}(u_{1},\cdots ,u_{p}|z)$ and $\Lambda (u,u_{1},\cdots ,u_{p}|z)$ being $c$
number functions. The vanishing of the so-called unwanted terms, ${\cal F}%
_{\alpha }=0$, is enforced in the usual procedure of the algebraic Bethe
Ansatz by finding the parameters $u_{1},...,u_{p}$. In this case the wave
vector $\Phi (u_{1},\cdots ,u_{p})$ becomes an eigenvector of the transfer
matrix with eigenvalue $\Lambda (u,u_{1},\cdots ,u_{p}|z)$. If we keep all
unwanted terms, i.e. ${\cal F}_{\alpha }\neq 0$, then the wave vector $\Phi $
in general satisfies the equation (\ref{int.4}), named in \cite{B} as
off-shell Bethe Ansatz equation ({\small OSBAE}).

There is a neat relationship between the wave vector satisfying the {\small %
OSBAE} (\ref{int.4}) and the vector-valued solutions of the {\small KZ}
equation (\ref{int.3}): The general vector valued solution of the {\small KZ}
equation for an arbitrary simple Lie algebra was found by Schechtman and
Varchenko \cite{SV}. It can be represented as a multiple contour integral

\begin{equation}
\Psi (z_{1},\ldots ,z_{N})=\oint \cdots \oint {\cal X}(u_{1},...,u_{p}|z)%
\phi (u_{1},...,u_{p}|z)du_{1}\cdots du_{p}.  \label{int.5}
\end{equation}%
The complex variables $z_{1},...,z_{N}$ of (\ref{int.5}) are related with
the disorder parameters of the {\small OSBAE} . The vector valued function $%
\phi (u_{1},...,u_{p}|z)$ is the semiclassical limit of the wave vector $%
\Phi (u_{1},...,u_{p}|z)$. In fact, it is the Bethe wave vector for Gaudin
magnets \cite{GA}, but \ "off -shell ". The Bethe Ansatz for the Gaudin
model was derived for any simple Lie algebra by Reshetikhin and Varchenko 
\cite{RV}. The scalar function ${\cal X}(u_{1},...,u_{p}|z)$ is constructed
from the semiclassical limit of the $\Lambda (u=z_{k};u_{1},...,u_{p}|z)$
and the ${\cal F}_{\alpha }(u_{1},\cdots ,u_{p}|z)$ functions. This
representation of the $N$-point correlation function shows a deep connection
between the inhomogeneous vertex models and the {\small WZNW \ }theory.

In this work we show that this idea also holds for the case of the
semiclassical limit that corresponds to the $sl(2|1)^{(2)}$ trigonometric $r$%
-matrix \cite{KS, BS}.

The paper is organized as follows. In Section $2$ we present the algebraic
Bethe Ansatz for the $sl(2|1)^{(2)}$ vertex model. Here the inhomogeneous
Bethe Ansatz is read from the homogeneous case previously derived for the $%
19 $-vertex models \cite{LI}. We also derive the off-shell Bethe Ansatz
equation for this vertex model. In Section $3$ , taking into account the
semiclassical limit of the results presented in the Section $2$, we describe
the algebraic structure of the corresponding Gaudin model. In Section $4$,
data of the off-shell Gaudin equation are used to construct solutions of the
trigonometric {\small KZ} equation. Conclusions are reserved for Section $5$.

\section{Inhomogeneous Algebraic Bethe Ansatz}

Consider $V=V_{0}\oplus V_{1}$ a $Z_{2}$-graded vector space where $0$ and $%
1 $ denote the even and odd parts respectively. The components of a linear
operator $A\overset{s}{\otimes }B$ in the graded tensor product space $V%
\overset{s}{\otimes }V$ result in matrix elements of the form 
\begin{equation}
(A\overset{s}{\otimes }B)_{\alpha \beta }^{\gamma \delta }=(-)^{p(\beta
)(p(\alpha )+p(\gamma ))}\ A_{\alpha \gamma }B_{\beta \delta }  \label{gra.1}
\end{equation}%
and the action of the permutation operator ${\cal P}$ on the vector $%
\left\vert \alpha \right\rangle \overset{s}{\otimes }\left\vert \beta
\right\rangle \in V\overset{s}{\otimes }V$ is given by 
\begin{equation}
{\cal P}\ \left\vert \alpha \right\rangle \overset{s}{\otimes }\left\vert
\beta \right\rangle =(-)^{p(\alpha )p(\beta )}\left\vert \beta \right\rangle 
\overset{s}{\otimes }\left\vert \alpha \right\rangle \Longrightarrow ({\cal P%
})_{\alpha \beta }^{\gamma \delta }=(-)^{p(\alpha )p(\beta )}\delta _{\alpha
\beta }\ \delta _{\gamma \delta },  \label{gra.2}
\end{equation}%
where $p(\alpha )=1\ (0)$ if $\left\vert \alpha \right\rangle $ is an odd
(even) element.

Besides ${\cal R}$ , it is usual to consider the matrix $R={\cal PR}$ which
satisfy 
\begin{equation}
R_{12}(u)R_{23}(u+v)R_{12}(v)=R_{23}(v)R_{12}(u+v)R_{23}(u).  \label{gra.2a}
\end{equation}

The regular solution of the graded {\small YB} equation of the $19$-vertex
model $sl(2|1)^{(2)}$ is given by \cite{KS, BS} 
\begin{equation}
R(u,\eta )=\left( 
\begin{array}{ccccccccccc}
x_{1} & 0 & 0 &  & 0 & 0 & 0 &  & 0 & 0 & 0 \\ 
0 & x_{5} & 0 &  & x_{2} & 0 & 0 &  & 0 & 0 & 0 \\ 
0 & 0 & x_{7} &  & 0 & x_{6} & 0 &  & x_{3} & 0 & 0 \\ 
&  &  &  &  &  &  &  &  &  &  \\ 
0 & x_{2} & 0 &  & x_{5} & 0 & 0 &  & 0 & 0 & 0 \\ 
0 & 0 & -x_{6} &  & 0 & -x_{4} & 0 &  & -x_{6} & 0 & 0 \\ 
0 & 0 & 0 &  & 0 & 0 & x_{5} &  & 0 & x_{2} & 0 \\ 
&  &  &  &  &  &  &  &  &  &  \\ 
0 & 0 & x_{3} &  & 0 & x_{6} & 0 &  & x_{7} & 0 & 0 \\ 
0 & 0 & 0 &  & 0 & 0 & x_{2} &  & 0 & x_{5} & 0 \\ 
0 & 0 & 0 &  & 0 & 0 & 0 &  & 0 & 0 & x_{1}%
\end{array}%
\right) ,  \label{gra.3}
\end{equation}%
where 
\begin{eqnarray}
x_{1}(u) &=&\cosh (u+\eta )\sinh (u+2\eta ),  \nonumber \\
x_{2}(u) &=&\sinh u\cosh (u+\eta ),  \nonumber \\
x_{3}(u) &=&\sinh u\cosh (u-\eta ),\quad  \nonumber \\
x_{5}(u) &=&\sinh 2\eta \cosh (u+\eta ),\qquad  \nonumber \\
x_{6}(u) &=&\sinh 2\eta \sinh u,\ \ \qquad  \nonumber \\
x_{7}(u) &=&\sinh 2\eta \cosh \eta \   \nonumber \\
x_{4}(u) &=&x_{2}(u)-x_{7}(u)  \label{gra.4}
\end{eqnarray}

Consider now an inhomogeneous vertex model, where to each vertex we
associate two parameters: a global spectral parameter $u$ and a disorder
parameter $z$. In this case, the vertex weight matrix ${\cal R}$ depends on $%
u-z$ and consequently the monodromy matrix defined bellow will be a function
of the disorder parameters $z_{i}$.

The graded quantum inverse scattering method is characterized by the
monodromy matrix $T(u|z)$ satisfying the equation 
\begin{equation}
R(u-v)\left[ T(u|z)\overset{s}{\otimes }T(v|z)\right] =\left[ T(v|z)\overset{%
s}{\otimes }T(u|z)\right] R(u-v),  \label{gra.5}
\end{equation}%
whose consistency is guaranteed by the graded version of the {\small YB}
equation (\ref{int.1}). $T(u|z)$ is a matrix in the space $V$ (usually
called auxiliary space) whose matrix elements are operators on the states of
the quantum system (which will also be in our work the space $V$). The
monodromy operator $T(u|z)$ is defined as an ordered product of local
operators ${\cal L}_{n}$ (Lax operator), on all sites of the lattice: 
\begin{equation}
T(u|z)={\cal L}_{N}(u-z_{N}){\cal L}_{N-1}(u-z_{N-1})\cdots {\cal L}%
_{1}(u-z_{1}).  \label{gra.6}
\end{equation}%
We normalize the Lax operator on the $n^{th}$ quantum space to be given by: 
\begin{eqnarray}
{\cal L}_{n} &=&\frac{1}{x_{2}}\left( 
\begin{array}{ccccccccccc}
x_{1} & 0 & 0 &  & 0 & 0 & 0 &  & 0 & 0 & 0 \\ 
0 & x_{2} & 0 &  & x_{5} & 0 & 0 &  & 0 & 0 & 0 \\ 
0 & 0 & x_{3} &  & 0 & x_{6} & 0 &  & x_{7} & 0 & 0 \\ 
&  &  &  &  &  &  &  &  &  &  \\ 
0 & x_{5} & 0 &  & x_{2} & 0 & 0 &  & 0 & 0 & 0 \\ 
0 & 0 & x_{6} &  & 0 & x_{4} & 0 &  & x_{6} & 0 & 0 \\ 
0 & 0 & 0 &  & 0 & 0 & x_{2} &  & 0 & x_{5} & 0 \\ 
&  &  &  &  &  &  &  &  &  &  \\ 
0 & 0 & x_{7} &  & 0 & x_{6} & 0 &  & x_{3} & 0 & 0 \\ 
0 & 0 & 0 &  & 0 & 0 & x_{5} &  & 0 & x_{2} & 0 \\ 
0 & 0 & 0 &  & 0 & 0 & 0 &  & 0 & 0 & x_{1}%
\end{array}%
\right)  \nonumber \\
&=&\left( 
\begin{array}{lll}
L_{11}^{(n)}(u-z_{n}) & L_{12}^{(n)}(u-z_{n}) & L_{13}^{(n)}(u-z_{n}) \\ 
L_{21}^{(n)}(u-z_{n}) & L_{22}^{(n)}(u-z_{n}) & L_{23}^{(n)}(u-z_{n}) \\ 
L_{31}^{(n)}(u-z_{n}) & L_{32}^{(n)}(u-z_{n}) & L_{33}^{(n)}(u-z_{n})%
\end{array}%
\right)  \label{gra.7}
\end{eqnarray}
$L_{\alpha \beta }^{(n)}(u),\ \alpha ,\beta =1,2,3$ are $3$ by $3$ matrices
acting on the $n^{th}$ site of the lattice, so the monodromy matrix has the
form 
\begin{equation}
T(u|z)=\left( 
\begin{array}{lll}
A_{1}(u|z) & B_{1}(u|z) & B_{2}(u|z) \\ 
C_{1}(u|z) & A_{2}(u|z) & B_{3}(u|z) \\ 
C_{2}(u|z) & C_{3}(u|z) & A_{3}(u|z)%
\end{array}%
\right) ,  \label{gra.8}
\end{equation}%
where 
\begin{eqnarray}
T_{ij}(u|z) &=&\sum_{k_{1},...,k_{N-1}=1}^{3}L_{ik_{1}}^{(N)}(u-z_{N})%
\overset{s}{\otimes }L_{k_{1}k_{2}}^{(N-1)}(u-z_{N-1})\overset{s}{\otimes }%
\cdots \overset{s}{\otimes }L_{k_{N-1}j}^{(1)}(u-z_{1}).  \nonumber \\
i,j &=&1,2,3.  \label{gra.9}
\end{eqnarray}

The vector in the quantum space of the monodromy matrix $T(u|z)$ that is
annihilated by the operators $T_{ij}(u|z)$, $i>j$ ($C_{i}(u|z)$ operators, $%
i=1,2,3$) and it is also an eigenvector for the operators $T_{ii}(u|z)$ ( $%
A_{i}(u|z)$ operators, $i=1,2,3$) is called a highest \ weight vector of the
monodromy matrix $T(u|z)$.

The transfer matrix $\tau (u|z)$ of the corresponding integrable spin model
is given by the supertrace of the monodromy matrix in the space $V$ 
\begin{equation}
\tau (u|z)=\sum_{i=1}^{3}(-1)^{p(i)}\
T_{ii}(u|z)=A_{1}(u|z)-A_{2}(u|z)+A_{3}(u|z).  \label{gra.10}
\end{equation}

The algebraic Bethe Ansatz solution for the inhomogeneous $sl(2|1)^{(2)}$
vertex model can be obtained from the homogeneous case \cite{LI}. The proper
modification is a local shift of the spectral parameter $u\rightarrow
u-z_{i} $.

Here we will define the functions used in our algebraic Bethe Ansatz:

\begin{eqnarray}
z(u) &=&\frac{x_{1}(u)}{x_{2}(u)}=\frac{\sinh (u+2\eta )}{\sinh u},\quad
y(u)=\frac{x_{3}(u)}{x_{6}(u)}=\frac{\cosh (u-\eta )}{\sinh 2\eta },\quad 
\nonumber \\
&&  \nonumber \\
\omega (u) &=&-\frac{x_{1}(u)x_{3}(u)}{x_{4}(u)x_{3}(u)-x_{6}(u)x_{6}(u)}=-%
\frac{\sinh (u+2\eta )\cosh (u-\eta )}{\sinh (u-2\eta )\cosh (u+\eta )}, 
\nonumber \\
&&  \nonumber \\
{\cal Z}(u_{k}-u_{j}) &=&\left\{ 
\begin{array}{c}
z(u_{k}-u_{j})\qquad \qquad \quad \quad {\rm if}\quad k>j \\ 
z(u_{k}-u_{j})\omega (u_{j}-u_{k})\quad \ {\rm if}\quad k<j%
\end{array}%
\right. .  \label{inh.1}
\end{eqnarray}%
We start defining the highest weight vector of the monodromy matrix $T(u|z)$
in a lattice of $N$ sites as the even (bosonic) completely unoccupied state 
\begin{equation}
\left\vert 0\right\rangle =\otimes _{a=1}^{N}\left( 
\begin{array}{c}
1 \\ 
0 \\ 
0%
\end{array}%
\right) _{a}.  \label{inh.2}
\end{equation}%
Using (\ref{gra.9}) we can compute the normalized action of the monodromy
matrix entries on this state 
\begin{eqnarray}
A_{i}(u|z)\left\vert 0\right\rangle &=&X_{i}(u|z)\left\vert 0\right\rangle
,\quad C_{i}(u|z)\left\vert 0\right\rangle =0,\quad B_{i}(u|z)\left\vert
0\right\rangle \neq \left\{ 0,\left\vert 0\right\rangle \right\} ,  \nonumber
\\
X_{i}(u|z) &=&\prod_{a=1}^{N}\frac{x_{i}(u-z_{a})}{x_{2}(u-z_{a})},\qquad
i=1,2,3.  \label{inh.3}
\end{eqnarray}

The Bethe vectors are defined as normal ordered states $\Psi
_{n}(u_{1},\cdots ,u_{n})$ which can be written with aid of a recurrence
formula \cite{TA}: 
\begin{eqnarray}
&&\left. \Psi _{n}(u_{1},...,u_{n}|z)=B_{1}(u_{1}|z)\Psi
_{n-1}(u_{2},...,u_{n}|z)\right.  \nonumber \\
&&  \nonumber \\
&&\left. -B_{2}(u_{1}|z)\sum_{j=2}^{n}\frac{X_{1}(u_{j}|z)}{y(u_{1}-u_{j})}%
\prod_{k=2,k\neq j}^{n}{\cal Z}(u_{k}-u_{j})\Psi _{n-2}(u_{2},...,\overset{%
\wedge }{u}_{j},...,u_{n}|z)\right. ,  \label{inh.5}
\end{eqnarray}
with the initial condition $\Psi _{0}=\left| 0\right\rangle ,\quad \Psi
_{1}(u_{1}|z)=B_{1}(u_{1}|z)\left| 0\right\rangle $. Here \ $\overset{\wedge 
}{u}_{j}$ denotes that the rapidity $u_{j}$ is absent: $\Psi (\overset{%
\wedge }{u}_{j}|z)=\Psi (u_{1},...,u_{j-1},u_{j+1},\cdots ,u_{n}|z) $.

The action of the transfer matrix $\tau (u|z)$ on these Bethe vectors gives
us the following off-shell Bethe Ansatz equation for the $sl(2|1)^{(2)}$
vertex model

\begin{equation}
\tau (u|z)\Psi _{n}(u_{1},...,u_{n}|z)=\Lambda _{n}\Psi
_{n}(u_{1},...,u_{n}|z)-\sum_{j=1}^{n}{\cal F}_{j}^{(n-1)}\Psi
_{(n-1)}^{j}+\sum_{j=2}^{n}\sum_{l=1}^{j-1}{\cal F}_{lj}^{(n-2)}\Psi
_{(n-2)}^{lj}.  \label{inh.4}
\end{equation}

We now briefly describe each term which appear in the right hand side of (%
\ref{inh.4}) ( for more details the reader can see \cite{LI}): in the first
term the Bethe vectors (\ref{inh.5}) are multiplied by $c$-numbers functions 
$\Lambda _{n}=\Lambda _{n}(u,u_{1},...,u_{n}|z)$ given by 
\begin{equation}
\Lambda
_{n}=X_{1}(u|z)\prod_{k=1}^{n}z(u_{k}-u)-(-)^{n}X_{2}(u|z)\prod_{k=1}^{n}%
\frac{z(u-u_{k})}{\omega (u-u_{k})}+X_{3}(u|z)\prod_{k=1}^{n}\frac{%
x_{2}(u-u_{k})}{x_{3}(u-u_{k})}.  \label{inh.6}
\end{equation}%
The second term is a sum of new vectors 
\begin{equation}
\Psi _{(n-1)}^{j}=\left( \frac{x_{5}(u_{j}-u)}{x_{2}(u_{j}-u)}B_{1}(u|z)+%
\frac{1}{y(u-u_{j})}B_{3}(u|z)\right) \Psi _{n-1}(\overset{\wedge }{u}_{j}),
\label{inh.7}
\end{equation}%
multiplied by scalar functions ${\cal F}_{j}^{(n-1)}$ given by 
\begin{equation}
{\cal F}_{j}^{(n-1)}=X_{1}(u_{j}|z)\prod_{k\neq j}^{n}{\cal Z}%
(u_{k}-u_{j})+(-)^{n}X_{2}(u_{j}|z)\prod_{k\neq j}^{n}{\cal Z}(u_{j}-u_{k}).
\label{inh.8}
\end{equation}%
Finally, the last term is a coupled sum of a third type of vector-valued
functions 
\begin{equation}
\Psi _{(n-2)}^{lj}=B_{2}(u|z)\Psi _{n-2}(\overset{\wedge }{u}_{l},\overset{%
\wedge }{u}_{j}),  \label{inh.9}
\end{equation}%
with coefficients 
\begin{eqnarray}
{\cal F}_{lj}^{(n-2)} &=&G_{lj}X_{1}(u_{l}|z)X_{1}(u_{j}|z)\prod_{k=1,k\neq
j,l}^{n}{\cal Z}(u_{k}-u_{l}){\cal Z}(u_{k}-u_{j})  \nonumber \\
&&-(-)^{n}Y_{lj}X_{1}(u_{l}|z)X_{2}(u_{j}|z)\prod_{k=1,k\neq j,l}^{n}{\cal Z}%
(u_{k}-u_{l}){\cal Z}(u_{j}-u_{k})  \nonumber \\
&&-(-)^{n}F_{lj}X_{1}(u_{j}|z)X_{2}(u_{l}|z)\prod_{k=1,k\neq j,l}^{n}{\cal Z}%
(u_{l}-u_{k}){\cal Z}(u_{k}-u_{j})  \nonumber \\
&&+H_{lj}X_{2}(u_{l}|z)X_{2}(u_{j}|z)\prod_{k=1,k\neq j,l}^{n}{\cal Z}%
(u_{j}-u_{k}){\cal Z}(u_{l}-u_{k}).  \label{inh.10}
\end{eqnarray}%
where $G_{lj}$ , $Y_{lj}$ , $F_{lj}$ and $H_{lj}$ are additional functions
defined by 
\begin{eqnarray}
G_{lj} &=&\frac{x_{7}(u_{l}-u)}{x_{3}(u_{l}-u)}\frac{1}{y(u_{l}-u_{j})}+%
\frac{z(u_{l}-u)}{\omega (u_{l}-u)}\frac{x_{5}(u_{j}-u)}{x_{2}(u_{j}-u)}%
\frac{1}{y(u-u_{l})},  \nonumber \\
&&  \nonumber \\
H_{lj} &=&\frac{x_{7}(u-u_{l})}{x_{3}(u-u_{l})}\frac{1}{y(u_{l}-u_{j})}-%
\frac{x_{5}(u-u_{l})}{x_{3}(u-u_{l})}\frac{1}{y(u-u_{j})},  \nonumber \\
&&  \nonumber \\
Y_{lj} &=&\frac{1}{y(u-u_{l})}\left\{ z(u-u_{l})\frac{x_{5}(u-u_{j})}{%
x_{2}(u-u_{j})}-\frac{y_{5}(u-u_{l})}{x_{2}(u-u_{l})}\frac{x_{5}(u_{l}-u_{j})%
}{x_{2}(u_{l}-u_{j})}\right\} ,  \nonumber \\
&&  \nonumber \\
F_{lj} &=&\frac{x_{5}(u-u_{l})}{x_{2}(u-u_{l})}\left\{ \frac{%
x_{5}(u_{l}-u_{j})}{x_{2}(u_{l}-u_{j})}\frac{1}{y(u-u_{l})}+\frac{z(u-u_{l})%
}{\omega (u-u_{l})}\frac{1}{y(u-u_{j})}\right.  \nonumber \\
&&\left. -\frac{x_{5}(u-u_{l})}{x_{2}(u-u_{l})}\frac{1}{y(u_{l}-u_{j})}%
\right\} .  \label{inh.11}
\end{eqnarray}%
In the usual Bethe Ansatz method, the next step consist in impose the
vanishing of the so-called unwanted terms of (\ref{inh.4}) in order to get
an eigenvalue problem for the transfer matrix.

We impose ${\cal F}_{j}^{(n-1)}=0$ and ${\cal F}_{lj}^{(n-2)}=0$ into (\ref%
{inh.4}) to recover the eigenvalue problem. This means that $\Psi
_{n}(u_{1},...,u_{n}|z)$ is an eigenstate of $\tau (u|z)$ with eigenvalue $%
\Lambda _{n}$ , provided the rapidities $u_{j}$ are solutions of the
inhomogeneous Bethe Ansatz equations 
\begin{eqnarray}
\prod_{a=1}^{N}z(u_{j}-z_{a}) &=&(-)^{n+1}\prod_{k=1,\ k\neq j}^{n}\frac{%
z(u_{j}-u_{k})}{z(u_{k}-u_{j})}\omega (u_{k}-u_{j}),  \nonumber \\
j &=&1,2,...,n.  \label{inh.12}
\end{eqnarray}

\section{Structure of the ${\bf sl(2|1)}^{(2)}$ Gaudin Model}

In this section we will consider the theory of the Gaudin models. To do this
we need to calculate the semiclassical limit of the results presented in the
previous section.

First, we recall that the superalgebra $sl(2|1)\simeq sl(1|2)$ is the ($N=2$%
) extended supersymmetric version of $sl(2)$ and contains four even
(bosonic) generators $H,W,X^{\pm }$ which form the Lie algebra $sl(2)\oplus
U(1)$ and four odd (fermionic) generators $V^{\pm },W^{\pm }$ whose
non-vanishing commutation relations read as \cite{MM, FSS}: 
\begin{eqnarray}
\lbrack H,X^{\pm }] &=&\pm X^{\pm },\qquad \quad \lbrack
X^{+},X^{-}]=2H,\qquad \quad \ \{V^{+},V^{-}\}=-\frac{1}{2}H.  \nonumber \\
\lbrack H,V^{\pm }] &=&\pm \frac{1}{2}V^{\pm },\qquad \ [X^{\pm },V^{\mp
}]=V^{\pm },\qquad \quad \ \{V^{\pm },V^{\pm }\}=\pm \frac{1}{2}X^{\pm }, 
\nonumber \\
\lbrack X^{\pm },V^{\pm }] &=&0,\ \qquad \qquad \ [X^{\pm },W^{\pm
}]=0,\qquad \quad \ \ \ \{V^{\pm },W^{\pm }\}=0,  \nonumber \\
\lbrack X^{\pm },W^{\mp }] &=&X^{\pm },\qquad \qquad \ [W,W^{\pm }]=\mp 
\frac{1}{2}X^{\pm },\qquad \ [W,X^{\pm }]=\frac{1}{2}W^{\pm },  \nonumber \\
\{V^{\pm },W^{\mp }\} &=&\pm \frac{1}{2}W^{\pm },\quad \ \{W^{+},W^{-}\}=%
\frac{1}{2}H,\quad \ \ \ \ \{W^{\pm },W^{\pm }\}=X^{\pm }.  \label{str.1}
\end{eqnarray}%
The quadratic Casimir operator is 
\begin{equation}
C_{2}=H^{2}-W^{2}+X^{-}X^{+}+2W^{-}W^{+}-2V^{-}V^{+},  \label{str.2}
\end{equation}%
and the elementary representation is three-dimensional, given by 
\begin{eqnarray}
W &=&\frac{1}{2}\left( 
\begin{array}{lll}
1 & 0 & 0 \\ 
0 & 2 & 0 \\ 
0 & 0 & 1%
\end{array}%
\right) ,\quad H=\frac{1}{2}\left( 
\begin{array}{lll}
1 & 0 & 0 \\ 
0 & 0 & 0 \\ 
0 & 0 & -1%
\end{array}%
\right) ,\ \quad X^{+}=\left( 
\begin{array}{lll}
0 & 0 & 1 \\ 
0 & 0 & 0 \\ 
0 & 0 & 0%
\end{array}%
\right) ,\qquad  \nonumber \\
&&  \nonumber \\
\ X^{-} &=&\left( 
\begin{array}{lll}
0 & 0 & 0 \\ 
0 & 0 & 0 \\ 
1 & 0 & 0%
\end{array}%
\right) ,\quad V^{+}=\frac{1}{2}\left( 
\begin{array}{lll}
0 & 1 & 0 \\ 
0 & 0 & 1 \\ 
0 & 0 & 0%
\end{array}%
\right) ,\ \quad V^{-}=\frac{1}{2}\left( 
\begin{array}{lll}
\ \ 0 & 0 & 0 \\ 
-1 & 0 & 0 \\ 
\ \ 0 & 1 & 0%
\end{array}%
\right) ,  \nonumber \\
&&  \nonumber \\
W^{+} &=&\frac{1}{2}\left( 
\begin{array}{lll}
0 & 1 & 0 \\ 
0 & 0 & -1 \\ 
0 & 0 & 0%
\end{array}%
\right) ,\ W^{-}=\frac{1}{2}\left( 
\begin{array}{lll}
0 & 0 & 0 \\ 
1 & 0 & 0 \\ 
0 & 1 & 0%
\end{array}%
\right)  \label{str.3}
\end{eqnarray}%
Here the basis is $\left\{ \left\vert 1\right\rangle ,\left\vert
0\right\rangle ,\left\vert -1\right\rangle \right\} $, where the first and
third vectors are considered as even and the second as odd, {\it i.e}., the
grading is {\small BFB}.

In order to expand the matrix elements of $T(u|z)$, up to an appropriate
order in $\eta $, we will start by expanding the Lax operator entries
defined in (\ref{gra.7}): 
\begin{eqnarray}
L_{11}^{(n)} &=&1+2\eta \frac{2-{\cal W}_{n}+{\cal H}_{n}\cosh (2u-2z_{n})}{%
\sinh (2u-2z_{n})}+4\eta ^{2}\left( \frac{1}{2}{\cal H}_{n}^{2}-\frac{1}{4}%
\frac{{\cal H}_{n}^{2}-{\cal H}_{n}}{\cosh (u-z_{n})^{2}}\right) +{\rm o}%
(\eta ^{3}),  \nonumber \\
&&  \nonumber \\
L_{22}^{(n)} &=&1+2\eta \frac{1-{\cal W}_{n}}{\sinh (2u-2z_{n})}+4\eta
^{2}\left( \frac{1}{2}\frac{1-{\cal H}_{n}^{2}}{\cosh (u-z_{n})^{2}}\right) +%
{\rm o}(\eta ^{3}),  \nonumber \\
&&  \nonumber \\
L_{33}^{(n)} &=&1+2\eta \frac{2-{\cal W}_{n}-{\cal H}_{n}\cosh (2u-2z_{n})}{%
\sinh (2u-2z_{n})}+4\eta ^{2}\left( \frac{1}{2}{\cal H}_{n}^{2}-\frac{1}{4}%
\frac{{\cal H}_{n}^{2}+{\cal H}_{n}}{\cosh (u-z_{n})^{2}}\right) +{\rm o}%
(\eta ^{3}).  \nonumber \\
&&  \label{gau.1}
\end{eqnarray}%
and for the elements out of the diagonal we have 
\begin{eqnarray}
L_{12}^{(n)} &=&2\eta \frac{{\rm e}^{u-z_{n}}{\cal W}_{a}^{-}-{\rm e}%
^{-u+z_{n}}{\cal V}_{a}^{-}}{\sinh (2u-2z_{n})}+{\rm o}(\eta ^{2}),\quad \
L_{21}^{(n)}=2\eta \frac{{\rm e}^{-u+z_{n}}{\cal W}_{a}^{+}+{\rm e}^{u-z_{n}}%
{\cal V}_{a}^{+}}{\sinh (2u-2z_{n})}+{\rm o}(\eta ^{2}),  \nonumber \\
&&  \nonumber \\
L_{23}^{(n)} &=&2\eta \frac{{\rm e}^{u-z_{n}}{\cal W}_{a}^{-}+{\rm e}%
^{-u+z_{n}}{\cal V}_{a}^{-}}{\sinh (2u-2z_{n})}+{\rm o}(\eta ^{2}),\quad
\quad L_{32}^{(n)}=2\eta \frac{{\rm e}^{u-z_{n}}{\cal V}_{a}^{+}-{\rm e}%
^{-u+z_{n}}{\cal W}_{a}^{+}}{\sinh (2u-2z_{n})}+{\rm o}(\eta ^{2}), 
\nonumber \\
&&  \nonumber \\
L_{13}^{(n)} &=&2\eta \ \frac{{\cal X}_{n}^{-}}{\sinh (2u-2z_{n})}+{\rm o}%
(\eta ^{2}),\quad \quad L_{31}^{(n)}=2\eta \ \frac{{\cal X}_{n}^{+}}{\sinh
(2u-2z_{n})}+{\rm o}(\eta ^{2}).  \label{gau.2}
\end{eqnarray}%
where ${\cal V}^{\pm }=2V^{\pm }$, ${\cal X}^{\pm }=2X^{\pm }$ , ${\cal W}%
^{\pm }=2W^{\pm }$, $\ {\cal H}=2H$ and ${\cal W}=2W$.

Substituting (\ref{gau.1}) and (\ref{gau.2}) into (\ref{gra.9}) we will get
the semiclassical expansion for the monodromy matrix entries.\ For the
diagonal entries we get 
\begin{equation}
A_{i}(u|z)=1+2\eta A_{i}^{(1)}(u|z)+4\eta ^{2}A_{i}^{(2)}(u|z)+{\rm o}(\eta
^{3}),\quad i=1,2,3.
\end{equation}%
where, the first order terms are given by%
\begin{eqnarray}
A_{1}^{(1)}(u|z) &=&\sum_{a=1}^{N}\frac{2-{\cal W}_{a}+{\cal H}_{a}\cosh
(2u-2z_{a})}{\sinh (2u-2z_{a})},\quad A_{2}^{(1)}(u|z)=\sum_{a=1}^{N}\frac{%
2-2{\cal W}_{a}}{\sinh (2u-2z_{a})},  \nonumber \\
A_{3}^{(1)}(u|z) &=&\sum_{a=1}^{N}\frac{2-{\cal W}_{a}-{\cal H}_{a}\cosh
(2u-2z_{a})}{\sinh (2u-2z_{a})}
\end{eqnarray}%
and the second order terms are%
\begin{eqnarray*}
A_{1}^{(2)}(u|z) &=&\sum_{a=1}^{N}(\frac{1}{2}{\cal H}_{a}^{2}-\frac{1}{4}%
\frac{{\cal H}_{a}^{2}-{\cal H}_{a}}{\cosh (2u-2z_{a})^{2}})+\sum_{a<b}\coth
(2u-2z_{a})\coth (2u-2z_{b}){\cal H}_{a}\overset{s}{\otimes }{\cal H}_{b} \\
&&+\sum_{a<b}\frac{(2-{\cal W}_{a})\overset{s}{\otimes }(2-{\cal W}_{b})}{%
\sinh (2u-2z_{a})\sinh (2u-2z_{b})}+\sum_{a<b}\frac{(2-{\cal W}_{a})\overset{%
s}{\otimes }{\cal H}_{b}\cosh (2u-2z_{b})}{\sinh (2u-2z_{a})\sinh (2u-2z_{b})%
} \\
&&+\sum_{a<b}\frac{\cosh (2u-2z_{a}){\cal H}_{a}\overset{s}{\otimes }(2-%
{\cal W}_{b})}{\sinh (2u-2z_{a})\sinh (2u-2z_{b})}+\sum_{a<b}\frac{{\cal X}%
_{a}^{-}\overset{s}{\otimes }{\cal X}_{b}^{+}}{\sinh (2u-2z_{a})\sinh
(2u-2z_{b})} \\
&&+\sum_{a<b}\frac{{\rm e}^{u-z_{a}}{\cal W}_{a}^{-}-{\rm e}^{-u+z_{a}}{\cal %
V}_{a}^{-}}{\sinh (2u-2z_{a})}\overset{s}{\otimes }\frac{{\rm e}^{-u-z_{b}}%
{\cal W}_{b}^{+}+{\rm e}^{u-z_{b}}{\cal V}_{b}^{+}}{\sinh (2u-2z_{b})},
\end{eqnarray*}%
\begin{eqnarray*}
A_{3}^{(2)}(u|z) &=&\sum_{a=1}^{N}(\frac{1}{2}{\cal H}_{a}^{2}-\frac{1}{4}%
\frac{{\cal H}_{a}^{2}+{\cal H}_{a}}{\cosh (2u-2z_{a})^{2}})+\sum_{a<b}\coth
(2u-2z_{a})\coth (2u-2z_{b}){\cal H}_{a}\overset{s}{\otimes }{\cal H}_{b} \\
&&+\sum_{a<b}\frac{(2-{\cal W}_{a})\overset{s}{\otimes }(2-{\cal W}_{b})}{%
\sinh (2u-2z_{a})\sinh (2u-2z_{b})}-\sum_{a<b}\frac{(2-{\cal W}_{a})\overset{%
s}{\otimes }{\cal H}_{b}\cosh (2u-2z_{b})}{\sinh (2u-2z_{a})\sinh (2u-2z_{b})%
} \\
&&-\sum_{a<b}\frac{\cosh (2u-2z_{a}){\cal H}_{a}\overset{s}{\otimes }(2-%
{\cal W}_{b})}{\sinh (2u-2z_{a})\sinh (2u-2z_{b})}+\sum_{a<b}\frac{{\cal X}%
_{a}^{+}\overset{s}{\otimes }{\cal X}_{b}^{-}}{\sinh (2u-2z_{a})\sinh
(2u-2z_{b})} \\
&&+\sum_{a<b}\left( \frac{{\rm e}^{u-z_{a}}{\cal V}_{a}^{+}-{\rm e}%
^{-u+z_{a}}{\cal W}_{a}^{+}}{\sinh (2u-2z_{a})}\right) \overset{s}{\otimes }%
\left( \frac{{\rm e}^{u-z_{b}}{\cal W}_{b}^{-}+{\rm e}^{-u+z_{b}}{\cal V}%
_{b}^{-}}{\sinh (2u-2z_{b})}\right) ,
\end{eqnarray*}
\begin{eqnarray}
A_{2}^{(2)}(u|z) &=&\sum_{a=1}^{N}\left( \frac{1}{2}\frac{1-{\cal H}_{a}^{2}%
}{\cosh (2u-2z_{a})^{2}}\right) +\sum_{a<b}\frac{(2-{\cal W}_{a})\overset{s}{%
\otimes }(2-{\cal W}_{b})}{\sinh (2u-2z_{a})\sinh (2u-2z_{b})}  \nonumber \\
&&+\sum_{a<b}\frac{{\rm e}^{u-z_{a}}{\cal W}_{a}^{+}+{\rm e}^{-u+z_{a}}{\cal %
V}_{a}^{+}}{\sinh (2u-2z_{a})}\overset{s}{\otimes }\frac{{\rm e}^{u-z_{b}}%
{\cal W}_{b}^{-}-{\rm e}^{-u+z_{b}}{\cal V}_{b}^{-}}{\sinh (2u-2z_{b})} 
\nonumber \\
&&+\sum_{a<b}\frac{{\rm e}^{u-z_{a}}{\cal W}_{a}^{-}+{\rm e}^{-u+z_{a}}{\cal %
V}_{a}^{-}}{\sinh (2u-2z_{a})}\overset{s}{\otimes }\frac{{\rm e}^{u-z_{b}}%
{\cal V}_{b}^{+}-{\rm e}^{-u+z_{b}}{\cal W}_{b}^{+}}{\sinh (2u-2z_{b})}.
\label{gau.3}
\end{eqnarray}%
For off-diagonal elements we only need to expand them up to the first order
in $\eta $%
\begin{eqnarray}
B_{1}(u|z) &=&2\eta \sum_{a=1}^{N}\frac{{\rm e}^{u-z_{a}}{\cal W}_{a}^{-}-%
{\rm e}^{-u+z_{a}}{\cal V}_{a}^{-}}{\sinh (2u-2z_{a})}+{\rm o}(\eta
^{2}),\quad  \nonumber \\
B_{2}(u|z) &=&2\eta \sum_{a=1}^{N}\frac{{\cal X}_{a}^{-}}{\sinh (2u-2z_{a})}+%
{\rm o}(\eta ^{2}),  \nonumber \\
B_{3}(u|z) &=&2\eta \sum_{a=1}^{N}\frac{{\rm e}^{u-z_{a}}{\cal W}_{a}^{-}+%
{\rm e}^{-u+z_{a}}{\cal V}_{a}^{-}}{\sinh (2u-2z_{a})}+{\rm o}(\eta ^{2}), 
\nonumber \\
C_{1}(u|z) &=&2\eta \sum_{a=1}^{N}\frac{{\rm e}^{-u+z_{a}}{\cal W}_{a}^{+}+%
{\rm e}^{u-z_{a}}{\cal V}_{a}^{+}}{\sinh (2u-2z_{a})}+{\rm o}(\eta ^{2}), 
\nonumber \\
C_{2}(u|z) &=&2\eta \sum_{a=1}^{N}\frac{{\cal X}_{a}^{+}}{\sinh (u-z_{a})}+%
{\rm o}(\eta ^{2})  \nonumber \\
C_{3}(u|z) &=&2\eta \sum_{a=1}^{N}\frac{{\rm e}^{u-z_{a}}{\cal V}_{a}^{+}-%
{\rm e}^{-u+z_{a}}{\cal W}_{a}^{+}}{\sinh (2u-2z_{a})}+{\rm o}(\eta ^{2}),
\label{gau.4}
\end{eqnarray}%
From these we have the following expansion for the transfer matrix (\ref%
{gra.10}): 
\begin{eqnarray}
\tau (u|z) &=&1+2\eta \sum_{a=1}^{N}\frac{2}{\sinh (2u-2z_{a})}+4\eta
^{2}\left\{ \sum_{a=1}^{N}({\cal H}_{a}^{2}-\frac{1}{2}\frac{1}{\cosh
(u-z_{a})^{2}})\right.  \nonumber \\
&&+\sum_{a<b}^{N}\frac{2}{\sinh (2u-2z_{a})\sinh (2u-2z_{b})}\!\!\left\{ 
{\cal H}_{a}\overset{s}{\otimes }{\cal H}_{b}\cosh (2u-2z_{a})\cosh
(2u-2z_{b})\right.  \nonumber \\
&&\left. -{\cal W}_{a}\overset{s}{\otimes }{\cal W}_{b}+\frac{1}{2}({\cal X}%
_{a}^{+}\overset{s}{\otimes }{\cal X}_{b}^{-}+{\cal X}_{a}^{-}\overset{s}{%
\otimes }{\cal X}_{b}^{+}+4)\right.  \nonumber \\
&&\left. +\left. {\rm e}^{-z_{a}+z_{b}}({\cal W}_{a}^{-}\overset{s}{\otimes }%
{\cal W}_{b}^{+}+{\cal V}_{a}^{+}\overset{s}{\otimes }{\cal V}_{b}^{-})-{\rm %
e}^{z_{a}-z_{b}}({\cal W}_{a}^{+}\overset{s}{\otimes }{\cal W}_{b}^{-}+{\cal %
V}_{a}^{-}\overset{s}{\otimes }{\cal V}_{b}^{+}\!\!)\right\} \right\} 
\nonumber \\
&\equiv &1+2\eta \tau ^{(1)}(u|z)+4\eta ^{2}\tau ^{(2)}(u|z)+{\rm o}(\eta
^{2}).  \label{gau.5}
\end{eqnarray}%
Now we will consider the second order term of $\tau (u|z)$: 
\begin{equation}
\tau ^{(2)}(u|z)=\sum_{a<b}^{N}{\cal G}_{ab}(u)+\sum_{a=1}^{N}({\cal H}%
_{a}^{2}-\frac{1}{2}\frac{1}{\cosh (u-z_{a})^{2}}).  \label{gau.5a}
\end{equation}%
which, with aid of the identity%
\begin{equation}
\frac{1}{\sinh (2u-2z_{a})\sinh (2u-2z_{b})}=\frac{1}{\sinh (2z_{a}-2z_{b})}(%
\frac{{\rm e}^{-2u+2z_{a}}}{\sinh (2u-2z_{a})}-\frac{{\rm e}^{-2u+2z_{b}}}{%
\sinh (2u-2z_{b})})  \label{gau.5b}
\end{equation}%
can be written in the form%
\begin{equation}
\tau ^{(2)}(u|z)=\sum_{a=1}^{N}\frac{{\cal G}_{a}(u)}{{\rm e}%
^{2u-2z_{a}}\sinh (2u-2z_{a})}+\sum_{a=1}^{N}({\cal H}_{a}^{2}-\frac{1}{2}%
\frac{1}{\cosh (u-z_{a})^{2}})  \label{gau.5c}
\end{equation}%
where 
\begin{eqnarray}
{\cal G}_{a}(u) &=&\sum_{b\neq a}\frac{2}{\sinh (2z_{a}-2z_{b})}\left\{
2+\cosh (2u-2z_{a})\cosh (2u-2z_{b}){\cal H}_{a}\overset{s}{\otimes }{\cal H}%
_{b}\right.  \nonumber \\
&&\left. -{\cal W}_{a}\overset{s}{\otimes }{\cal W}_{b}+\frac{1}{2}\left( 
{\cal X}_{a}^{+}\overset{s}{\otimes }{\cal X}_{b}^{-}+{\cal X}_{a}^{-}%
\overset{s}{\otimes }{\cal X}_{b}^{+}\right) \right.  \nonumber \\
&&\left. +\left. {\rm e}^{-z_{a}+z_{b}}\left( {\cal W}_{a}^{-}\overset{s}{%
\otimes }{\cal W}_{b}^{+}+{\cal V}_{a}^{+}\overset{s}{\otimes }{\cal V}%
_{b}^{-}\right) -{\rm e}^{z_{a}-z_{b}}\left( {\cal W}_{a}^{+}\overset{s}{%
\otimes }{\cal W}_{b}^{-}+{\cal V}_{a}^{-}\overset{s}{\otimes }{\cal V}%
_{b}^{+}\right) \!\!\right\} \right\}  \nonumber \\
&&  \label{gau.5d}
\end{eqnarray}%
Here we observe that \ ${\cal G}_{a}(u)$ is nothing but the sum of
semiclassical \ trigonometric $r$-matrices. This fact follows from the
construction of the semiclassical $r$-matrices out of the quadratic Casimir:%
\begin{equation}
C_{2}=H^{2}-W^{2}+\frac{1}{2}%
(X^{-}X^{+}+X^{+}X^{-})+(W^{-}W^{+}+V^{+}V^{-})-(W^{+}W^{-}+V^{-}V^{+}).
\label{gau.6}
\end{equation}

The Gaudin Hamiltonians are defined as the residue of $\tau (u|z)$ at the
point $u=z_{a}$. This results in $N$ non-local Hamiltonians%
\begin{eqnarray}
G_{a} &=&\sum_{b\neq a}^{N}\frac{1}{\sinh (2z_{a}-2z_{b})}\left\{ 2+\cosh
(2z_{a}-2z_{b}){\cal H}_{a}\overset{s}{\otimes }{\cal H}_{b}\right. 
\nonumber \\
&&-{\cal W}_{a}\overset{s}{\otimes }{\cal W}_{b}+\frac{1}{2}\left( {\cal X}%
_{a}^{+}\overset{s}{\otimes }{\cal X}_{b}^{-}+{\cal X}_{a}^{-}\overset{s}{%
\otimes }{\cal X}_{b}^{+}\right)  \nonumber \\
&&+\left. {\rm e}^{-z_{a}+z_{b}}({\cal W}_{a}^{-}\overset{s}{\otimes }{\cal W%
}_{b}^{+}+{\cal V}_{a}^{+}\overset{s}{\otimes }{\cal V}_{b}^{-})-{\rm e}%
^{z_{a}-z_{b}}({\cal W}_{a}^{+}\overset{s}{\otimes }{\cal W}_{b}^{-}+{\cal V}%
_{a}^{-}\overset{s}{\otimes }{\cal V}_{b}^{+})\right\} ,  \nonumber \\
a &=&1,2,...,N.  \label{gau.9}
\end{eqnarray}%
satisfying 
\begin{equation}
\sum_{a=1}^{N}G_{a}=0,\quad \frac{\partial G_{a}}{\partial z_{b}}=\frac{%
\partial G_{b}}{\partial z_{a}},\quad \left[ G_{a},G_{b}\right] =0,\qquad
\forall a,b.  \label{gau.10}
\end{equation}

In the next section we will use the data of the Bethe Ansatz presented in
the previous section in order to find the exact spectrum and eigenvectors
for each of these $N-1$ independent Hamiltonians.

We complete this section deriving \ the $sl(2|1)^{(2)}$ Gaudin algebra from
the semiclassical limit of the fundamental commutation relation (\ref{gra.5}%
): The semiclassical expansions of $T$ and $R$ can be written in the
following form 
\begin{equation}
T(u|z)=1+2\eta \lbrack l(u|z)+\beta (u|z)]+o(\eta ^{2}),\quad R(u)={\cal P}%
\left[ 1+2\eta \lbrack r(u)+\beta (u)]+o(\eta ^{2})\right] .  \label{gau.11}
\end{equation}%
where%
\begin{equation}
\beta (u|z)=\sum_{a=1}^{N}\frac{2}{\sinh (2u-2z_{a})}.  \label{gau.11b}
\end{equation}

Using (\ref{gau.3}--\ref{gau.4}) one can see that the \ \textquotedblright
classical $l$-operator \textquotedblright\ has the form%
\begin{equation}
l(u|z)=\left( 
\begin{array}{ccccc}
{\cal W}(u|z)+{\cal H}(u|z) &  & {\cal W}^{-}(u|z)-{\cal V}^{-}(u|z) &  & 
{\cal X}^{-}(u|z) \\ 
&  &  &  &  \\ 
{\cal W}^{+}(u|z)+{\cal V}^{+}(u|z) &  & 2{\cal W}(u|z) &  & {\cal W}%
^{-}(u|z)+{\cal V}^{-}(u|z) \\ 
&  &  &  &  \\ 
{\cal X}^{+}(u|z) &  & -{\cal W}^{+}(u|z)+{\cal V}^{+}(u|z) &  & {\cal W}%
(u|z)-{\cal H}(u|z)%
\end{array}%
\right)  \label{gau.12}
\end{equation}%
where 
\begin{eqnarray}
{\cal H}(u|z) &=&\sum_{a=1}^{N}\coth (2u-2z_{a}){\cal H}_{a},\qquad {\cal W}%
(u|z)=\sum_{a=1}^{N}\frac{-{\cal W}_{a}}{\sinh (2u-2z_{a})}  \nonumber \\
{\cal W}^{-}(u|z) &=&\sum_{a=1}^{N}\frac{{\rm e}^{u-z_{a}}}{\sinh (2u-2z_{a})%
}{\cal W}_{a}^{-},\qquad {\cal W}^{+}(u|z)=\sum_{a=1}^{N}\frac{{\rm e}%
^{-u+z_{a}}}{\sinh (2u-2z_{a})}{\cal W}_{a}^{+},  \nonumber \\
{\cal V}^{-}(u|z) &=&\sum_{a=1}^{N}\frac{{\rm e}^{-u+z_{a}}}{\sinh
(2u-2z_{a})}{\cal V}_{a}^{-},\qquad {\cal V}^{+}(u|z)=\sum_{a=1}^{N}\frac{%
{\rm e}^{u-z_{a}}}{\sinh (2u-2z_{a})}{\cal V}_{a}^{+},  \nonumber \\
{\cal X}^{-}(u|z) &=&\sum_{a=1}^{N}\frac{{\cal X}_{a}^{-}}{\sinh (2u-2z_{a})}%
,\qquad {\cal X}^{+}(u|z)=\sum_{a=1}^{N}\frac{{\cal X}_{a}^{+}}{\sinh
(2u-2z_{a})}.  \label{gau.13}
\end{eqnarray}%
The corresponding semiclassical $r$-matrix has the form 
\begin{eqnarray}
r(u) &=&\frac{1}{\sinh 2u}\left\{ \cosh 2u\ {\cal H}\overset{s}{\otimes }%
{\cal H-W}\overset{s}{\otimes }{\cal W}+\frac{1}{2}({\cal X}^{+}\overset{s}{%
\otimes }{\cal X}^{-}+{\cal X}^{-}\overset{s}{\otimes }{\cal X}^{+})\right. 
\nonumber \\
&&+{\rm e}^{u}\ ({\cal W}^{-}\overset{s}{\otimes }{\cal W}^{+}+{\cal V}^{+}%
\overset{s}{\otimes }{\cal V}^{-})-{\rm e}^{-u}\ ({\cal W}^{-}\overset{s}{%
\otimes }{\cal W}^{+}+{\cal V}^{-}\overset{s}{\otimes }{\cal V}^{+}).
\label{gau.14}
\end{eqnarray}%
Here we notice that\ (\ref{gau.14}) is equivalent to the $r$-matrix
constructed out of the quadratic Casimir (\ref{gau.6}) in a standard way 
\cite{Ku2}.

Substituting (\ref{gau.14}) and (\ref{gau.12}) into (\ref{gra.5}), we have 
\begin{eqnarray}
&&{\cal P}l(u|z)\overset{s}{\otimes }l(v|z)+{\cal P}r(u-v)\left[ l(u|z)%
\overset{s}{\otimes }1+1\overset{s}{\otimes }l(v|z)\right]  \nonumber \\
&=&l(v|z)\overset{s}{\otimes }l(u|z){\cal P}+\left[ l(v|z)\overset{s}{%
\otimes }1+1\overset{s}{\otimes }l(u|z)\right] {\cal P}r(u-v),
\label{gau.16}
\end{eqnarray}%
whose consistence is guaranteed by the graded classical {\small YB} equation
(\ref{int.2}).

From (\ref{gau.16}) we can derive (anti-)commutation relations between the
matrix elements of $l(u|z)$. This gives us the defining relations of the $%
sl(2|1)^{(2)}$ Gaudin algebra :%
\begin{eqnarray}
\lbrack {\cal H}(u|z),{\cal H}(v|z)] &=&0,\quad \lbrack {\cal W}(u|z),{\cal H%
}(v|z)]=0,\quad \lbrack {\cal W}(u|z),{\cal W}(v|z)]=0,  \nonumber \\
\quad \lbrack {\cal X}^{\pm }(u|z),{\cal W}(v|z)] &=&0,\quad \lbrack {\cal X}%
^{\pm }(u|z),{\cal X}^{\pm }(v|z)]=0,\quad \lbrack {\cal X}^{\pm }(u|z),%
{\cal W}^{\pm }(v|z)]=0,  \nonumber \\
\lbrack {\cal X}^{\pm }(u|z),{\cal V}^{\pm }(v|z)] &=&0,\quad \{{\cal V}%
^{\pm }(u|z),{\cal W}^{\pm }(v|z)\}=0,  \nonumber \\
\lbrack {\cal X}^{\mp }(u|z),{\cal X}^{\pm }(v|z)] &=&\pm 4\frac{{\cal H}%
(u|z)-{\cal H}(v|z)}{\sinh (2u-2v)},  \nonumber \\
\lbrack {\cal W}^{\pm }(u|z),{\cal W}(v|z)] &=&\frac{{\cal V}^{\pm }(u|z)-%
{\rm e}^{\pm (-u+v)}{\cal V}^{\pm }(v|z)}{\sinh (2u-2v)}  \nonumber \\
\lbrack {\cal V}^{\pm }(u|z),{\cal X}^{\mp }(v|z)] &=&2\frac{{\cal V}^{\mp
}(u|z)-{\rm e}^{\pm (u-v)}{\cal V}^{\mp }(v|z)}{\sinh (2u-2v)},  \nonumber \\
\lbrack {\cal W}^{\pm }(u|z),{\cal X}^{\mp }(v|z)] &=&2\frac{{\cal W}^{\mp
}(u|z)-{\rm e}^{\pm (-u+v)}{\cal W}^{\mp }(v|z)}{\sinh (2u-2v)},  \nonumber
\\
\lbrack {\cal V}^{\pm }(u|z),{\cal W}(v|z)] &=&\frac{{\cal W}^{\pm }(u|z)-%
{\rm e}^{\pm (u-v)}{\cal W}^{\pm }(v|z)}{\sinh (2u-2v)},  \nonumber \\
\lbrack {\cal X}^{\pm }(u|z),{\cal H}(v|z)] &=&\pm 2\frac{\cosh (2u-2v){\cal %
X}^{\pm }(u|z)-{\cal X}^{\pm }(v|z)}{\sinh (2u-2v)},  \nonumber \\
\lbrack {\cal V}^{\pm }(u|z),{\cal H}(v|z)] &=&\pm \frac{\cosh (2u-2v){\cal V%
}^{\pm }(u|z)-{\rm e}^{\pm (u-v)}{\cal V}^{\pm }(v|z)}{\sinh (2u-2v)}, 
\nonumber \\
\lbrack {\cal W}^{\pm }(u|z),{\cal H}(v|z)] &=&\pm \frac{\cosh (2u-2v){\cal W%
}^{\pm }(u|z)-{\rm e}^{\pm (-u+v)}{\cal W}^{\pm }(v|z)}{\sinh (2u-2v)}, 
\nonumber \\
\{{\cal V}^{\pm }(u|z),{\cal V}^{\mp }(v|z)\} &=&\frac{{\rm e}^{\pm
(u-v)}\left( {\cal H}(u|z)-{\cal H}(v|z)\right) }{\sinh (2u-2v)},  \nonumber
\\
\{{\cal W}^{\pm }(u|z),{\cal W}^{\mp }(v|z)\} &=&-\frac{{\rm e}^{\pm
(-u+v)}\left( {\cal H}(u|z)-{\cal H}(v|z)\right) }{\sinh (2u-2v)},  \nonumber
\\
\{{\cal V}^{\pm }(u|z),{\cal W}^{\mp }(v|z)\} &=&\pm \frac{{\rm e}^{\pm
(-u+v)}{\cal W}(u|z)-{\rm e}^{\pm (u-v)}{\cal W}(v|z)}{\sinh (2u-2v)}, 
\nonumber \\
\{{\cal W}^{\pm }(u|z),{\cal W}^{\pm }(v|z)\} &=&\pm \frac{{\rm e}^{\pm
(u-v)}{\cal X}^{\pm }(u|z)-{\rm e}^{\pm (-u+v)}{\cal X}^{\pm }(v|z)}{\sinh
(2u-2v)},  \nonumber \\
\{{\cal V}^{\pm }(u|z),{\cal V}^{\pm }(v|z)\} &=&\mp \frac{{\rm e}^{\pm
(-u+v)}{\cal X}^{\pm }(u|z)-{\rm e}^{\pm (u-v)}{\cal X}^{\pm }(v|z)}{\sinh
(2u-2v)}.  \label{gau.17}
\end{eqnarray}%
A direct consequence of these relations is the commutativity of $\tau
^{(2)}(u|z)$ 
\begin{equation}
\lbrack \tau ^{(2)}(u|z),\tau ^{(2)}(v|z)]=0,\qquad \forall u,v
\label{gau.18}
\end{equation}%
from which the commutativity of the Gaudin Hamiltonians $G_{a}$ follows
immediately.

\section{Off-shell Gaudin Equation}

In order to get the semiclassical limit of the {\small OSBAE} (\ref{inh.4})
we first consider the semiclassical expansions of the Bethe vectors defined
in (\ref{inh.5}), (\ref{inh.7}) and (\ref{inh.9}): 
\begin{eqnarray}
\Psi _{n}(u_{1},...,u_{n}|z) &=&(2\eta )^{n}\Phi _{n}(u_{1},...,u_{n}|z)+%
{\rm o}(\eta ^{n+1}),  \nonumber \\
&&  \nonumber \\
\Psi _{(n-1)}^{j} &=&-2(2\eta )^{n+1}\frac{\left[ {\cal W}^{-}(u|z){\rm e}%
^{-u+u_{j}}-{\cal V}^{-}(u|z){\rm e}^{u-u_{j}}\right] }{\sinh (2u-2u_{j})}%
\Phi _{n-1}(\overset{\wedge }{u}_{j}|z)+{\rm o}(\eta ^{n+2}),  \nonumber \\
&&  \nonumber \\
\Psi _{(n-2)}^{lj} &=&(2\eta )^{n-1}{\cal X}^{-}(u|z)\Phi _{n-2}(\overset{%
\wedge }{u}_{l},\overset{\wedge }{u}_{j}|z)+{\rm o}(\eta ^{n}),
\label{off.1}
\end{eqnarray}%
where 
\begin{eqnarray}
\Phi _{n}(u_{1},...,u_{n}|z) &=&\left[ {\cal W}^{-}(u_{1}|z)-{\cal V}%
^{-}(u_{1}|z)\right] \Phi _{n-1}(u_{2},...,u_{n}|z)  \nonumber \\
&&-{\cal X}^{-}(u_{1}|z)\sum_{j=2}^{n}\frac{(-)^{j}}{\cosh (u_{1}-u_{j})}%
\Phi _{n-2}(u_{2},\overset{\wedge }{u}_{j}\!,u_{n}|z),  \label{off.2}
\end{eqnarray}%
with $\Phi _{0}=\left\vert 0\right\rangle $ and $\Phi _{1}(u_{1}|z)=\left[ 
{\cal W}^{-}(u_{1}|z)-{\cal V}^{-}(u_{1}|z)\right] \Phi _{0}$.

The corresponding expansions of the $c$-number functions presented in the 
{\small OSBAE are:} (\ref{inh.4}) 
\begin{eqnarray}
\Lambda _{n} &=&1+2\eta \Lambda _{n}^{(1)}+4\eta ^{2}\Lambda _{n}^{(2)}+{\rm %
o}(\eta ^{3}),  \label{off.3} \\
&&  \nonumber \\
{\cal F}_{j}^{(n-1)} &=&2\eta (-)^{j+1}\ f_{j}^{(n-1)}+{\rm o}(\eta ^{2}),
\label{off.4} \\
&&  \nonumber \\
{\cal F}_{lj}^{(n-2)} &=&2(2\eta )^{3}\frac{(-)^{l+j}}{\cosh (u_{l}-u_{j})}%
\left\{ \frac{f_{l}^{(n-1)}}{\sinh (2u-2u_{l})}-\frac{f_{j}^{(n-1)}}{\sinh
(2u-2u_{j})}\right\} +{\rm o}(\eta ^{4}),  \nonumber \\
&&  \label{off.5}
\end{eqnarray}%
\ where%
\begin{equation}
\Lambda _{n}^{(1)}=\sum_{a=1}^{N}\frac{2}{\sinh (2u-2z_{a})}  \label{off.5a}
\end{equation}%
\begin{eqnarray}
\Lambda _{n}^{(2)} &=&N+n-\frac{1}{2}\sum_{a=1}^{N}\frac{1}{\cosh
(u-z_{a})^{2}}  \nonumber \\
&&-\sum_{a=1}^{N}\sum_{j=1}^{n}\left\{ \coth (u-z_{a})\coth (u-u_{j})+\tanh
(u-z_{a})\tanh (u-u_{j})\right\}  \nonumber \\
&&+\sum_{a<b}^{N}\left\{ \coth (u-z_{a})\coth (u-z_{b})+\tanh (u-z_{a})\tanh
(u-z_{b})\right\}  \nonumber \\
&&+\sum_{j<k}^{n}\left\{ \coth (u-u_{j})\coth (u-u_{k})+\tanh (u-u_{j})\tanh
(u-u_{k})\right.  \nonumber \\
&&+\left. \frac{4}{\sinh (2u-2u_{j})\sinh (2u-2u_{k})}\right\}  \label{off.6}
\end{eqnarray}%
and 
\begin{equation}
f_{j}^{(n-1)}=\sum_{a=1}^{N}\coth (u_{j}-z_{a})-\sum_{k\neq j}^{n}\tanh
(u_{j}-u_{k}).  \label{off.7}
\end{equation}

Substituting these expressions into the (\ref{inh.4}) and comparing the
coefficients of the terms $2(2\eta )^{n+2}$ we get the first non-trivial
consequence for the semiclassical limit of the \ {\small OSBAE}: 
\begin{equation}
\tau ^{(2)}(u|z)\ \Phi _{n}(u_{1},...,u_{n}|z)=\Lambda _{n}^{(2)}\ \Phi
_{n}(u_{1},...,u_{n}|z)-\sum_{j=1}^{n}(-)^{j}\ \frac{2f_{j}^{(n-1)}\Theta
_{(n-1)}^{j}}{\sinh (2u-2u_{j})}.  \label{off.8}
\end{equation}%
Note that in this limit the contributions from $\Psi _{(n-1)}^{j}$ and $\Psi
_{(n-2)}^{lj}$ are combined to give a new vector valued function 
\begin{eqnarray}
\Theta _{(n-1)}^{j} &=&\left[ {\cal W}^{-}(u|z){\rm e}^{-u+u_{j}}-{\cal V}%
^{-}(u|z){\rm e}^{u-u_{j}}\right] \ \Phi _{n-1}(\ \overset{\wedge }{u}_{j}|z)
\nonumber \\
&&-{\cal X}^{-}(u|z)\sum_{k=1,\ k\neq j}^{n}\frac{(-)^{k^{^{\prime }}}}{%
\cosh (u_{j}-u_{k})}\ \Phi _{n-2}(\ \overset{\wedge }{u}_{j},\overset{\wedge 
}{u}_{k}|z),  \label{off.9}
\end{eqnarray}%
where $k^{^{\prime }}=k+1\ \ $for$\quad k<j$ \ and $k^{^{\prime }}=k\ $\ for$%
\quad k>j$.

Finally, we take the residue of (\ref{off.8}) at the point $u=z_{a}$ to get
the off-shell Gaudin equation: 
\begin{eqnarray}
G_{a}\Phi _{n}(u_{1},...,u_{n}|z) &=&g_{a}\Phi
_{n}(u_{1},...,u_{n}|z)+\sum_{l=1}^{n}(-)^{l}\frac{2f_{l}^{(n-1)}\phi
_{(n-1)}^{l}}{\sinh (2u_{l}-2z_{a})},  \nonumber \\
a &=&1,2,...,N  \label{off.11}
\end{eqnarray}%
where $g_{a}$ is the residue of $\Lambda _{n}^{(2)}$ 
\begin{equation}
g_{a}={\rm res}_{u=z_{a}}\Lambda _{n}^{(2)}=\sum_{b\neq a}^{N}\coth
(z_{a}-z_{b})-\sum_{l=1}^{n}\coth (z_{a}-u_{l}),  \label{off.12}
\end{equation}%
and $\phi _{(n-1)}^{l}$ is the residue of $\Theta _{(n-1)}^{l}$ 
\begin{eqnarray}
\phi _{(n-1)}^{j} &=&{\rm res}_{u=z_{a}}\Theta _{(n-1)}^{j}  \nonumber \\
&=&\frac{1}{2}({\cal W}_{a}^{-}{\rm e}^{u_{j}-z_{a}}-{\cal V}_{a}^{-}{\rm e}%
^{-u_{j}+z_{a}})\Phi _{n-1}(\overset{\wedge }{u}_{j}|z)-\frac{1}{2}{\cal X}%
_{a}^{-}\sum_{k\neq j}^{n}\frac{(-)^{k^{\prime }}}{\cosh (u_{j}-u_{k})}\Phi
_{n-2}(\overset{\wedge }{u}_{k},\overset{\wedge }{u}_{j}|z).  \label{off.13}
\end{eqnarray}

The equation (\ref{off.11}) allows us solve one of the main problem of the
Gaudin model, {\it i.e.}, the determination of the eigenvalues and
eigenvectors of the commuting Hamiltonians $G_{a}$ (\ref{gau.6}): $g_{a}$ is
the eigenvalue of $G_{a}$ with eigenfunction $\Phi _{n}$ provided $u_{l}$
are solutions of the following equations $f_{j}^{(n-1)}=0$, {\it i.e}.: 
\begin{equation}
\sum_{k\neq j}^{n}\tanh (u_{j}-u_{k})=\sum_{a=1}^{N}\coth
(u_{j}-z_{a}),\quad j=1,2,...,n.  \label{off.14}
\end{equation}%
Moreover, as we will see in the next section, the off-shell Gaudin equation (%
\ref{off.11}) provides solutions for the differential equations known as 
{\small KZ} equations.

\section{Knizhnik-Zamolodchickov equation}

The {\small KZ} differential equation 
\begin{equation}
\kappa \frac{\partial \Psi (z)}{\partial z_{i}}=G_{i}(z)\Psi (z),
\label{kz.1}
\end{equation}%
appeared first as a \ holonomic system of differential equations of
conformal blocks in a {\small WZW} models of conformal field theory. Here $%
\Psi (z)$ is a function with values in the tensor product $V_{1}\otimes
\cdots \otimes V_{N}$ of representations of a simple Lie algebra, $\kappa
=k+g$ , where $k$ is the central charge of the associated Kac-Moody algebra,
and $g$ is the dual Coxeter number of the simple Lie algebra.

One of the remarkable properties of the {\small KZ} system is that the
coefficient functions $G_{i}(z)$ commute and that the form $\omega
=\sum_{i}G_{i}(z)dz_{i}$ is closed \cite{RV}: 
\begin{equation}
\frac{\partial G_{j}}{\partial z_{i}}=\frac{\partial G_{i}}{\partial z_{j}}%
,\qquad \left[ G_{i},G_{j}\right] =0.  \label{kz.2}
\end{equation}
Indeed, it was indicated in \cite{RV} that the equations (\ref{kz.2}) are
not just a flatness condition for the form $\omega $ but that the {\small KZ}
connection is actually a commutative family of connections.

In this section we will identify $G_{i}$ with our $sl(2|1)^{(2)}$ Gaudin
Hamiltonians $G_{a}$ derived in the previous section%
\begin{eqnarray}
G_{a} &=&\sum_{b\neq a}^{N}\frac{1}{\sinh (2z_{a}-2z_{b})}\left\{ \cosh
(2z_{a}-2z_{b}){\cal H}_{a}\overset{s}{\otimes }{\cal H}_{b}\right. 
\nonumber \\
&&-{\cal W}_{a}\overset{s}{\otimes }{\cal W}_{b}+\frac{1}{2}\left( {\cal X}%
_{a}^{+}\overset{s}{\otimes }{\cal X}_{b}^{-}+{\cal X}_{a}^{-}\overset{s}{%
\otimes }{\cal X}_{b}^{+}\right)  \nonumber \\
&&+\left. {\rm e}^{-z_{a}+z_{b}}({\cal W}_{a}^{-}\overset{s}{\otimes }{\cal W%
}_{b}^{+}+{\cal V}_{a}^{+}\overset{s}{\otimes }{\cal V}_{b}^{-})-{\rm e}%
^{z_{a}-z_{b}}({\cal W}_{a}^{+}\overset{s}{\otimes }{\cal W}_{b}^{-}+{\cal V}%
_{a}^{-}\overset{s}{\otimes }{\cal V}_{b}^{+})\right\} ,  \nonumber \\
a &=&1,2,...,N.
\end{eqnarray}
and show that the corresponding differential equations (\ref{kz.1}) can be
solved via the off-shell Bethe Ansatz method.

Let us now define the vector-valued function $\Psi (z_{1},...,z_{N})$
through multiple contour integrals of the Bethe vectors (\ref{off.2}) 
\begin{equation}
\Psi (z_{1},...,z_{N})=\oint \cdots \oint {\cal X}(u|z)\Phi
_{n}(u|z)du_{1}...du_{n},  \label{kz.3}
\end{equation}
where ${\cal X}$ $(u|z)={\cal X}$ $(u_{1},...,u_{n},z_{1},...,z_{N})$ is a
scalar function which in this stage is still undefined.

We assume that $\Psi (z_{1},...,z_{N})$ is a solution of the equations 
\begin{equation}
\kappa \frac{\partial \Psi (z_{1},...,z_{N})}{\partial z_{a}}=G_{a}\Psi
(z_{1},...,z_{N}),\quad a=1,2,...,N  \label{kz.4}
\end{equation}
where $G_{a}$ are the Gaudin Hamiltonians (\ref{gau.6}) and $\kappa $ is a
constant.

Substituting (\ref{kz.3}) into (\ref{kz.4}) we have 
\begin{equation}
\kappa \frac{\partial \Psi (z_{1},...,z_{N})}{\partial z_{a}}=\oint \left\{
\kappa \frac{\partial {\cal X}(u|z)}{\partial z_{a}}\Phi _{n}(u|z)+\kappa 
{\cal X}(u|z)\frac{\partial \Phi _{n}(u|z)}{\partial z_{a}}\right\} du,
\label{kz.5}
\end{equation}
where we are using a compact notation $\oint \left\{ \circ \right\} du=\oint
\ldots \oint \left\{ \circ \right\} $\ $du_{1}\cdots du_{n}.$

Using the Gaudin algebra (\ref{gau.17}) one can derive the following
non-trivial identity 
\begin{equation}
\frac{\partial \Phi _{n}}{\partial z_{a}}=-\sum_{l=1}^{n}(-)^{l}\frac{%
\partial }{\partial u_{l}}\left( \frac{2\phi _{(n-1)}^{l}}{\sinh
(2u_{l}-2z_{a})}\right) ,  \label{kz.6}
\end{equation}%
which allows us write (\ref{kz.5}) in the form 
\begin{eqnarray}
\kappa \frac{\partial \Psi }{\partial z_{a}} &=&\oint \left\{ \kappa \frac{%
\partial {\cal X}(u|z)}{\partial z_{a}}\Phi
_{n}(u|z)+\sum_{l=1}^{n}(-)^{l}\kappa \frac{\partial {\cal X}(u|z)}{\partial
u_{l}}\left( \frac{2\phi _{(n-1)}^{l}}{\sinh (2u_{l}-2z_{a})}\right)
\right\} du  \nonumber \\
&&-\kappa \sum_{l=1}^{n}(-)^{l}\oint \frac{\partial }{\partial u_{l}}\left( 
{\cal X}(u|z)\frac{2\phi _{(n-1)}^{l}}{\sinh (2u_{l}-2z_{a})}\right) du.
\label{kz.7}
\end{eqnarray}%
It is evident that the last term of (\ref{kz.7}) vanishes, because the
contours are closed. Moreover, if the scalar function ${\cal X}(u|z)$
satisfies the following differential equations 
\begin{equation}
\kappa \frac{\partial {\cal X}(u|z)}{\partial z_{a}}=g_{a}{\cal X}%
(u|z),\qquad \kappa \frac{\partial {\cal X}(u|z)}{\partial u_{j}}%
=f_{j}^{(n-1)}{\cal X}(u|z),  \label{kz.8}
\end{equation}%
we are recovering the off-shell Gaudin equation (\ref{off.11}) from the
first term in (\ref{kz.7}).

Taking into account the definition of the scalar functions \ $f_{j}^{(n-1)}$(%
\ref{off.7}) and $g_{a}$ (\ref{off.12}), we can see that the consistency
condition of the system (\ref{kz.8}) is insured by the zero curvature
conditions $\partial f_{j}^{(n-1)}/\partial z_{a}=\partial g_{a}/\partial
u_{j}$. Moreover, the solution of (\ref{kz.8}) is easily obtained 
\begin{equation}
{\cal X}(u|z)=\prod_{a<b}^{N}\sinh (z_{a}-z_{b})^{1/\kappa
}\prod_{j<k}^{n}\cosh (u_{j}-u_{k})^{1/\kappa
}\prod_{a=1}^{N}\prod_{j=1}^{n}\sinh (z_{a}-u_{j})^{-1/\kappa }.
\label{kz.9}
\end{equation}%
This function determines the monodromy of $\Psi (z_{1},...,z_{N})$ as
solution of the trigonometric {\small KZ} equation (\ref{kz.4}) and these
results are in agreement with the Schechtman-Varchenko construction for
multiple contour integral as solutions of the {\small KZ} equation in an
arbitrary simple Lie algebra \cite{SV}.

\section{Conclusion}

In this paper a graded $19$-vertex model was used to generalize previous
rational vertex models results connecting the Gaudin magnet models to the
semiclassical off-shell Bethe Ansatz of these vertex models.

Using the semiclassical limit of the transfer matrix of the vertex model we
derived the trigonometric $sl(2|1)^{(2)}$ Gaudin Hamiltonians. The reduction
of the off-shell Gaudin equation \ to an eigenvalue equation gives us the
exact spectra and eigenvectors for these Gaudin magnets. Data of the
off-shell Gaudin equation were used to show that a hypergeometric type
integral (\ref{kz.3}) is solution of the trigonometric {\small KZ}
differential equation. \ 

In fact, this method had already been used with success to constructing
solutions of trigonometric {\small KZ} equations \cite{B2, CH, LU} and
elliptic {\small KZ}-Bernard equations \cite{B3}, for the six-vertex model
and eight-vertex model, respectively.

\vspace{1cm}{}

{\bf Acknowledgment:} This work was supported in part by Funda\c{c}\~{a}o de
Amparo \`{a} Pesquisa do Estado de S\~{a}o Paulo--FAPESP--Brasil, by
Conselho Nacional de Desenvol\-{}vimento--CNPq--Brasil and by Coordena\c{c}%
\~{a}o de Aperfei\c{c}oamento de Pessoal de N\'{\i}vel
Superior--CAPES-Brasil.

\end{document}